\documentclass[aps,prl,preprint,groupedaddress]{revtex4}
\begin{document}

\title{Rough surface scattering in many-mode conducting
channels:\\ gradient versus amplitude scattering}

\author{F.~M.~Izrailev$^{1}$, N.~M.~Makarov$^{2}$, and
M.~Rend\'on$^{3}$}
\affiliation{ $^{1}$Instituto de F\'{\i}sica, Universidad
Aut\'{o}noma de Puebla,
Apartado Postal J-48, Puebla, Pue., 72570, M\'{e}xico\\
$^{2}$Instituto de Ciencias, Universidad Aut\'{o}noma de Puebla,
Priv. 17 Norte No 3417, Col. San Miguel Hue\-yo\-tlipan, Puebla,
Pue., 72050, M\'{e}xico\\
$^{3}$Facultad de Ciencias de la Electr\'onica, Universidad
Aut\'{o}noma de Puebla, Puebla, Pue., 72570, M\'{e}xico }

\begin{abstract}
We study the effect of surface scattering on transport properties
in many-mode conducting channels (electron waveguides). Assuming a strong
roughness of the surface profiles, we show that there are two independent control
parameters that determine statistical properties of the
scattering. The first parameter is the ratio of the amplitude of
the roughness to the transverse width of the waveguide. The second
one, which is typically omitted, is determined by the mean
value of the derivative of the profile. This parameter may be large,
thus leading to specific properties of scattering. Our results may
be used in experimental realizations of the surface scattering of
electron waves, as well as for other applications (e.g., for
optical and microwave waveguides).
\end{abstract}

\maketitle
%%%%%%%%%%%%%%%%%%%%%%%%%%%%%%%%%%%%%%%%%%%%%%%%%%%%%%%%%%%%%%%%%%%%
Recent numerical studies of quasi-1D disordered systems
\cite{recent,SFYM99} have revealed principal difference between
surface and bulk scattering (see, e.g., discussion and references
in \cite{LFL98}). Specifically, it was found that transport
properties of quasi-1D waveguides with rough surfaces essentially
depend on many characteristic lengths, in contrast to the bulk
scattering where one-parametric scaling is determined by the ratio
of the localization length to the lengthwise size of samples. This fact is
due to a non-isotropic character of surface scattering in the
``channel space''. In particular, the transmission coefficient
smoothly decreases with an increase of the angle of incoming
waves, see details in \cite{SFYM99,IM03}. Below we present an
analytical treatment of the electron/wave scattering in waveguides with
rough surfaces, paying main attention to the interplay between
``amplitude'' and ``gradient'' scattering mechanisms \cite{leon,MT98}.

Our model under consideration is a plane waveguide (or conducting
quasi-one-dimensional wire) of average width $d$, that stretches
along the $x$-axis. The lower boundary of the waveguide is defined
as $z=\sigma\xi_1(x)$, while the upper boundary has the profile
$z=d+\sigma\xi_2(x)$. Therefore, the width $d(x)$ of the waveguide
is $d(x)=d+\sigma[\xi_2(x)-\xi_1(x)]$ with $\langle d(x)\rangle
=d$. For random functions $\xi_1(x)$ and $\xi_2(x)$ we assume
$\langle\xi_1(x)\rangle=\langle\xi_2(x)\rangle=0$ and
$\langle\xi_1^2(x)\rangle = \langle\xi_2^2(x)\rangle = 1$, so that $\sigma$
takes the meaning of the root-mean-square roughness height.
Here the
angular brackets stand for the averaging over $x$ for specific
realizations of profiles $\xi_{1,2}(x)$ (or over different
realizations of $\xi_{1,2}(x)$). In what follows we consider three
cases that reveal generic characteristics of the surface
scattering: (A) only the upper profile is rough, $ \xi_1(x)=0$ and
$\xi_2(x)=\xi(x)$; (B) two profiles are asymmetrical,
$\xi_1(x)=\xi_2(x)=\xi(x)$, in respect to the central line
$z=d/2$; (C) two profiles are symmetrical, $
-\xi_1(x)=\xi_2(x)=\xi(x)$. Note, however, that our approach is
valid for any profiles $\xi_1(x)$ and $\xi_2(x)$.

The method we use is based on the coordinate transformation that
makes both boundaries flat, (see for example, \cite{MT98,flat}),

\begin{equation}\label{Gen-xCoorTr}
x_{new}=x_{old}=x, \qquad
z_{new}=\frac{[z_{old}-\sigma\xi_1(x)]d}{d(x)}.
\end{equation}

Let us start with the case (A) when one surface is flat and the
other has the roughness that is assumed to be defined by the
Gaussian random function $\xi(x)$ with the binary correlator
$\langle\xi(x)\xi(x')\rangle={\cal W}(x-x')$. The latter is
normalized to its maximal value, ${\cal W}(0)=1$, and supposed to
decrease on a characteristic scale $R_c$. Since ${\cal W}(x)$ is
an even function of $x$, its Fourier transform
$W(k_x)=\int_{-\infty}^{\infty}dx\,\exp(-ik_xx)\,{\cal W}(x)$ is
even, real and non-negative function of the lengthwise
wave number $k_x$.
The roughness power
spectrum $W(k_x)$ has a maximum  $W(0)\sim R_c$ at $k_x=0$, and
decreases on the scale $R_c^{-1}$ as $|k_x|$ increases.

In order to solve the scattering problem we employed the method of
the Green's function ${\cal G}(x,x';z,z')$ for which the equation
has the form,

\begin{equation} \label{1RB-Geq1}
\left(\frac{\partial^2}{\partial x^2}+ \frac{\partial^2}{\partial
z^2}+k^2\right){\cal G}(x,x';z,z') =\delta(x-x')\delta(z-z'),
\end{equation}
with the boundary conditions ${\cal G}(x,x';z=0,z')={\cal
G}(x,x';z=d(x),z')=0$.  The wave number $k$ is equal to $\omega/c$
for a classical scalar wave of frequency $\omega$, and is the
Fermi wave number for electrons in the isotropic Fermi-liquid
model.

After transformation to new variables the equation for the canonically
conjugated Green's function gets the following form (below we use notation $z$
instead of $z_{new}$),

\begin{eqnarray}\label{1RB-GFP4}
&&\Bigg(\frac{\partial^2}{\partial x^2}+
\frac{\partial^2}{\partial z^2} +k^2\Bigg) \,{\cal G}(x,x';z,z')
-\Bigg\{\bigg[1-\frac{d^2}{d^2(x)}\bigg]
\frac{\partial^2}{\partial z^2}+ \frac{\sigma{\hat{\cal
U}}(x)}{d(x)}\left[\frac{1}{2}+ z\frac{\partial}{\partial
z}\right]
\nonumber\cr\\
&& -\frac{\sigma^2{\xi'}^2(x)}{d^2(x)}
\bigg[\frac{3}{4}+3z\,\frac{\partial}{\partial z}+
z^2\frac{\partial^2}{\partial z^2}\bigg]\Bigg\}\, {\cal
G}(x,x';z,z') =\delta(x-x')\delta(z-z'),
\end{eqnarray}
with flat-boundary conditions, ${\cal G}(x,x';z=0,z')={\cal
G}(x,x';z=d,z')=0$. Here the operator ${\hat{\cal U}}(x)$ is
defined by

\begin{equation}\label{U-def}
{\hat{\cal U}}(x)=\xi'(x)\frac{\partial}{\partial x}+
\frac{\partial}{\partial x}\,\xi'(x)=\xi''(x)+
2\xi'(x)\frac{\partial}{\partial x}.
\end{equation}
We underline that Eq.~(\ref{1RB-GFP4}) is {\it exact} and valid
for any profile $\xi(x)$. It contains a term (in braces)
that plays the role of an effective potential. This potential
depends not only on the profile $\xi(x)$
({\it amplitude scattering}), but also on its first and second
derivatives $\xi'(x)$ and $\xi''(x)$ ({\it gradient scattering}).
This very fact demonstrates a highly non-trivial role of surface
scattering.

To proceed, we assume that the surface roughness is small in
height, $\sigma\ll d$, but can have any value of its slope
($\sigma$ and $R_c$ can be in arbitrary relation). The
small-height approximation is common in surface scattering
theories that are based on an appropriate perturbative approach
(see for example, Ref.~\cite{BFb79}). Using this approach, we
obtained the general expression for the inverse attenuation length
$L_n$ (or, mean free path) of the {\it n-}th conducting subchannel,

\begin{equation}\label{1RB-Ln-sum}
\frac{1}{L_n}=\frac{1}{L^{(1)}_n}+\frac{1}{L^{(2)}_n},
\end{equation}
which is represented as a sum of two terms for a better
understanding of the role of amplitude and gradient scattering.
The first attenuation length $L^{(1)}_n$ reads as
\begin{equation} \label{1RB-Ln1}
\frac{1}{L^{(1)}_n}=\sigma^2\frac{(\pi
n/d)^2}{k_nd}\,\sum_{n'=1}^{N_d} \frac{(\pi n'/d)^2}{k_{n'}d}
\left[ W(k_n+k_{n'})+W(k_n-k_{n'})\right],
\end{equation}
where $k_n=\sqrt{k^2 - (\pi n/d)^2}$, $n=1, 2, 3,\ldots, N_d$,
and $N_d$ is the total number of conducting subchannels.
Here the {\it diagonal term} is formed by the {\it amplitude
mechanism} of surface scattering while the {\it off-diagonal
terms} result from the {\it gradient scattering}. The above expression
exactly coincides with that obtained many years ago by different
methods (see, e.g., Ref.~\cite{BFb79}). The second attenuation
length $L^{(2)}_n$ can be represented in the form,
\begin{equation}\label{1RB-LnG-mode}
\frac{1}{L^{(2)}_n}=\sum_{n'=1}^{N_d}\frac{1}{L^{(2)}_{nn'}}=\frac{1}{L^{(2)}_{nn}}
+\sum_{n'\neq n}^{N_d} \frac{1}{L^{(2)}_{nn'}},
\end{equation}
where the diagonal term
\begin{equation}
\frac{1}{L^{(2)}_{nn}}=\frac{\sigma^4}{2}\frac{(\pi
n/d)^4}{k_n^2}\,\left[\frac{1}{3}+\frac{1}{(2\pi n)^2}\right]^2
\left[T(2k_n)+T(0)\right], \label{1RB-LnnG}
\end{equation}
with $T(k_x)=\int_{-\infty}^{\infty}dx\exp{(-ik_xx)}{{\cal
W}''}^2(x)$ controls the electron/wave scattering {\it inside} the
subchannel ({\it intramode scattering}). The off-diagonal partial
attenuation length $L^{(2)}_{n\neq n'}$ that describes the {\it
intermode scattering} (from $n$-th subchannel to $n'\neq n$ one), is

\begin{equation}
\frac{1}{L^{(2)}_{nn'}}=\frac{8\sigma^4}{\pi^4}\frac{(\pi
n/d)^2}{k_n}\,\frac{(\pi n'/d)^2}{k_{n'}}\,
\frac{(n^2+n'^2)^2}{(n^2-n'^2)^4}
\left[T(k_n+k_{n'})+T(k_n-k_{n'})\right]. \label{1RB-Lnn'2}
\end{equation}

To the best of our knowledge, in the previous studies of a surface
scattering the second term in Eq.~(\ref{1RB-Ln-sum}), i.e.
$1/L^{(2)}_{n}$, never was taken into account. In this relation, we should
emphasize the principal importance of this term. In spite of that $1/L^{(2)}_{n}$
is proportional to $\sigma^4 $, it can prevail over $1/L^{(1)}_{n}$
even in the small roughness regime $\sigma \ll d$. Indeed, both attenuation
lengths, $1/L^{(1)}_n$ and $1/L^{(2)}_n$, depend on $R_c$ via
the substantially different functions: the roughness-height power
spectrum $W(k_x)$ and the square-gradient power spectrum $T(k_x)$, respectively.

For asymmetric profiles (case (B)) the total width of a waveguide
is constant, $d(x)=d$. As a result, the scattering is due to the
gradient terms only. Using the above approach one can obtain,

\begin{equation}\label{Ln}
L_n^{-1}=\sum_{n'=1}^{N_d}L_{nn'}^{-1}.
\end{equation}
Remarkably, in this case the diagonal and off-diagonal terms in
Eq.~(\ref{Ln}) are rather distinct. Specifically, the diagonal
term is proportional to $\sigma^4$,

\begin{equation}\label{CCB-Lnn}
\frac{1}{L_{nn}}=\frac{\sigma^4}{2}\frac{(\pi n/d)^4}{k_n^2}
\left[T(2k_n)+T(0)\right],
\end{equation}
in comparison with off-diagonal terms, which are proportional to
$\sigma^2$,

\begin{equation}
\frac{1}{L_{nn'}}=4\sigma^2\frac{(\pi n/d)^2}{k_nd}\frac{(\pi
n'/d)^2}{k_{n'}d}\,\sin^4 \Big[\frac{\pi(n-n')}{2}\Big]
[W(k_n+k_{n'})+W(k_n-k_{n'})].
\label{CCB-Lnn'}
\end{equation}
From this equation one can see that due to specific symmetry of
the two surface profiles, transitions between subchannels with even
difference $n-n'$ are forbidden (corresponding partial scattering
lengths diverge). Therefore, only transitions between odd and even
subchannels are allowed.

For symmetric profiles (case (C)) the surface scattering is caused
by both amplitude and gradient mechanisms. The diagonal term in
Eq.~(\ref{Ln}) has the form,

\begin{equation}
\frac{1}{L_{nn}}=4\sigma^2\frac{(\pi n/d)^4}{(k_nd)^2} \left[
W(2k_n)+W(0)\right] +\frac{\sigma^4}{2}\frac{(\pi n/d)^4}{k_n^2}\,
\left[\frac{1}{3}+\frac{1}{(\pi n)^2}\right]^2
\left[T(2k_n)+T(0)\right].
\label{SSB-Lnn}
\end{equation}
According to our analysis, the term which is proportional to
$\sigma^2$ is due to the amplitude scattering, and the second term
($\propto \sigma^4$) results from the gradient scattering. Note that
in a single-mode waveguide with $N_d=1$ the sum over $n'$ in
Eq.~(\ref{Ln}) contains only one term with $n'=n=1$. In this case
the backscattering length $L_{11}^{(b)}$ which enters into
Eqs.~(\ref{CCB-Lnn}) and (\ref{SSB-Lnn}) is in accordance with
that obtained in Refs.~\cite{MT98,MT01}.

The off-diagonal partial attenuation length $L_{n\neq n'}$
(intermode scattering) is due to the gradient scattering only,

\begin{eqnarray}
&&\frac{1}{L_{nn'}}=4\sigma^2\frac{(\pi n/d)^2}{k_nd} \frac{(\pi
n'/d)^2}{k_{n'}d}\,\cos^4 \Big[ \frac{\pi(n-n')}{2} \Big]
[W(k_n+k_{n'})+W(k_n-k_{n'})] \nonumber \cr\\
&& +\frac{32\sigma^4}{\pi^4}\frac{(\pi n/d)^2}{k_n}\, \frac{(\pi
n'/d)^2}{k_{n'}}\,\frac{(n^2+n'^2)^2}{(n^2-n'^2)^4}
\cos^4 \Big[ \frac{\pi(n-n')}{2} \Big]
[T(k_n+k_{n'})+T(k_n-k_{n'})]. \nonumber \cr\\
\end{eqnarray}
The effect of absence of transitions between some subchannels arises
in this case, as well as in the case with asymmetric profiles (case (B)).
However, in contrast to the former, now there are no transitions between
the subchannels with odd difference of their indexes $n-n'$. Thus, only
transitions between even subchannels and between odd subchannels are
permitted.

In conclusion, we have studied the role of amplitude and gradient
scattering in quasi-1D waveguides with rough surfaces. Our results
for the models with different symmetries between upper and lower
profiles demonstrate a principal difference for these two
mechanisms of scattering.

%\begin{acknowledgement}
This research was supported by the
Universidad Aut\'onoma de Puebla (BUAP, M\'exico) under the grant
II-104G04.
%\end{acknowledgement}

\end{document}